\documentclass[preprint,prd,showpacs,amsmath,amssymb,amsthm,nofootinbib]{revtex4}
\usepackage[mathscr]{euscript}
\usepackage{bm}
\usepackage{graphicx}
\usepackage{subfigure}
\usepackage[colorlinks=true,linkcolor=red]{hyperref}
\newcommand{\bea}{\begin{eqnarray}}
\newcommand{\eea}{\end{eqnarray}}
\newcommand{\beq}{\begin{equation}}
\newcommand{\eeq}{\end{equation}}

\def\/{\over}

\begin{document}

\title{Stability of Einstein static universe in gravity theory with a non-minimal derivative coupling}
\author{Qihong Huang$^{1,2}$,  Puxun Wu$^{1,3}$ 
 and Hongwei Yu$^{1}$\footnote{Corresponding author: hwyu@hunnu.edu.cn}  }
\affiliation{$^1$ Department of Physics and Synergetic Innovation Center for Quantum Effects and Applications, Hunan Normal University, Changsha, Hunan 410081, China \\
$^2$School of Physics and Electronic Science, Zunyi Normal College, Zunyi 563006, China\\
$^3$Center for High Energy Physics, Peking University, Beijing 100080, China }

\begin{abstract}
The emergent mechanism provides a possible way to resolve the big bang singularity problem by  assuming that our universe originates from  the Einstein static (ES) state. Thus,  the existence of a stable ES solution  becomes a very crucial prerequisite for the emergent scenario.  In this paper, we study the stability of an ES universe in gravity theory with a non-minimal coupling between the kinetic term of a scalar field and the Einstein tensor. We find that  the ES solution is stable under both scalar and tensor perturbations when the model parameters satisfy certain conditions, which indicates that   the big bang singularity  can be avoided successfully by the emergent mechanism in  the non-minimally kinetic coupled gravity.
\end{abstract}


\maketitle

\section{Introduction}

Although the standard cosmological model achieves  great success, it still suffers from several theoretical problems. The attempt to resolve theses problems  leads to the invention of the inflation theory~\cite{Staro1980, Guth1981, Linde1982}, which  settles successfully  most of the problems in the standard cosmological model, but leaves the big bang singularity problem open. To avoid this problem, some theories, such as the pre-big bang~\cite{Gasperini2003}, the cyclic scenario~\cite{Khoury2001} and the emergent scenario~\cite{Ellis2004},  have been proposed. The emergent scenario,  proposed by Ellis $et$ $al.$, in the framework of general relativity~\cite{Ellis2004}, assumes  that the universe  originates from an  Einstein static (ES) state rather than a big bang singularity. So, it requires that the universe can stay in an ES state past eternally, and exit this static state naturally and then evolve to a subsequent inflationary era. Apparently,  a very crucial prerequisite for the emergent scenario is the existence of a stable ES solution under various perturbations, such as quantum fluctuations. However, in the framework of general relativity, the emergent mechanism  is not as successful  as one expected in avoiding the big bang singularity since there is no stable ES solution in a Friedmann universe with a scalar field  minimally coupled with gravity~\cite{Barrow2003}.

A natural generalization of the minimally coupled gravity is to assume a non-minimal coupling between the scalar field and the curvature, which can be generated naturally when quantum corrections are considered and is essential for the renormalizability of the scalar field theory in curved space. This non-minimally coupled scalar field has been suggested to be responsible for both the early cosmic inflation~\cite{Abbott} and the present accelerated expansion~\cite{Sahni}. If the coupling is a general  function of the scalar field, the resulting theory is called the scalar-tensor theory~\cite{Bergmann}. The popular modified gravity, $f(R)$ gravity~\cite{Staro}, can be cast into a special form of the Brans-Dicke theory, which is a particular example of the scalar-tensor theory~\cite{Brans},  with a potential for the effective scalar-field degree of freedom.  Let us also note that  a pioneering inflation model  was  constructed in a $f(R)$ theory by Starobinsky~\cite{Staro1980},   which allows a graceful exit from inflation to the subsequent radiation dominated stage and produces a very good fit to existing CMB observational data~\cite{Planck}.

It has been found that in $f(R)$ theory the inhomogeneous scalar perturbations break the stability of the ES solution which is stable under homogeneous scalar perturbations~\cite{ Barrow1983,Seahra2009}.  Recently, Miao et. al~\cite{Miao2016} found that there is no stable ES solution when scalar perturbations and tensor ones are considered together in the scalar-tensor theory of gravity with a normal perfect fluid, such as radiation or pressureless matter. It is worthy to note that the stability of ES solutions  have  also been analyzed in some other theories~\cite{Huang2014, Li2017, Wu2011, Mulryne2005, HuangQ2015, Zhang2016,Zhang2010, HuangH2015,Bohmer2009, Wu2010,Bohmer2015,Atazadeh2015,HuangH2014,Bohmer2013,Tawfik2016}.

Except for the coupling between the scalar field and the curvature, there are many other coupling, such as the coupling between the kinetic term of the scalar field and the Einstein tensor.   This non-minimal derivative coupling has been discussed extensively  in cosmology. For example, it can 
provide an inflationary mechanism~\cite{Amendola1993, Capozziello1999, Granda2011, Yang2016, Darabi2015}, explain  both a quasi-de Sitter phase and an exit from it without any fine-tuned potential~\cite{Sush2009}, and  behave as a dark matter~\cite{Gao2010} or a dark energy~\cite{Granda2010, Staro2016}. Recently, the stability of  ES solutions  in the non-minimally derivative coupled gravity have been  studied in~\cite{Atazadeh2015}. However, in~\cite{Atazadeh2015} only a very special case of $\dot{\phi}=0$ is considered, which is not a general result derived from the conditions of  static state solution, where $\dot{\phi}=d\phi/dt$ with $\phi$  and $t$ being the scalar field and the cosmic time, respectively.  In addition, only the homogeneous scalar perturbations and tensor perturbations are considered in~\cite{Atazadeh2015}.   Thus, for the   non-minimally kinetic  coupled gravity, it is unclear whether the ES solution remain to be  stable against inhomogeneous scalar perturbations and  what the effect  the special condition $\dot{\phi}=0$ has on the stable regions, and this motivates us to the present work.

The paper is organized as follows. In Section 2, we give the field equations of gravity theory with a non-minimal derivative coupling and the ES solution. In Section 3, we analyze the stability of ES solution under tensor perturbations. In Section 4, the homogeneous and inhomogeneous scalar perturbations are considered. Finally, our main conclusions are presented in Section 5. Throughout this paper, unless specified, we adopt the metric signature ($-, +, +, +$). Latin indices run from 0 to 3 and the Einstein convention is assumed for repeated indices.

\section{The field equations and Einstein static solution}

The action of the  non-minimally derivative coupled gravity has the form~\cite{Amendola1993,Germani2010}
\bea\label{action}
S=\int d^{4}x \sqrt{-g}\Big[\frac{R}{8\pi G}-(g_{\mu\nu}+\kappa G_{\mu\nu})\nabla^{\mu}\phi\nabla^{\nu}\phi-2V(\phi)\Big]+S_{m},
\eea
where $R$ is the Ricci curvature scalar, $G$ is the Newtonian gravitational constant, $g_{\mu\nu}$ is the metric tensor with $g$ being its trace, $G_{\mu\nu}$ is the Einstein tensor, $V(\phi)$ is the potential of the scalar field $\phi$, $\kappa$ stands for the coupling parameter with dimension of (length)$^2$, and $S_{m}$ represents the action of a perfect fluid.

Varying the action~(\ref{action}) with respect to the metric tensor $g_{\mu\nu}$ and the scalar field $\phi$, respectively, one can obtain two independent equations:
\bea\label{Einstein}
G_{\mu\nu}=8\pi G[T_{\mu\nu}^{(m)}+T_{\mu\nu}^{(\phi)}+\kappa\vartheta_{\mu\nu}]
\eea
and
\bea\label{scalar}
(g^{\mu\nu}+\kappa G^{\mu\nu})\nabla_{\mu}\nabla_{\nu}\phi=V_{,\phi},
\eea
where $V_{,\phi}=\frac{dV}{d\phi}$, $T_{\mu\nu}^{(m)}$ is the energy-momentum tensor of the perfect fluid, and
\bea
&&T_{\mu\nu}^{(\phi)}=\nabla_{\mu}\phi\nabla_{\nu}\phi-\frac{1}{2}g_{\mu\nu}(\nabla\phi)^{2}-g_{\mu\nu}V,\\
&&\vartheta_{\mu\nu}=-\frac{1}{2}\nabla_{\mu}\phi\nabla_{\nu}\phi R+2\nabla_{\alpha}\phi \nabla_{(\mu}\phi R_{\nu)}^{\alpha}+\nabla^{\alpha}\phi\nabla^{\beta}\phi R_{\mu\alpha\nu\beta}
+\nabla_{\mu}\nabla^{\alpha}\phi\nabla_{\nu}\nabla_{\mu}\phi-\nabla_{\mu}\nabla_{\nu}\phi\square\phi \nonumber\\
&&\qquad\quad-\frac{1}{2}(\nabla\phi)^{2}G_{\mu\nu}+g_{\mu\nu}\bigg[-\frac{1}{2}\nabla^{\alpha}\nabla^{\beta}\phi\nabla_{\alpha}\nabla_{\beta}\phi+\frac{1}{2}(\square\phi)^{2}-\nabla_{\alpha}\phi\nabla_{\beta}\phi R^{\alpha\beta}\bigg].
\eea
Here, $(\nabla\phi)^{2}=\nabla_{\alpha}\phi\nabla^{\alpha}\phi$ and $\square\phi=\nabla_{\alpha}\nabla^{\alpha}\phi$.

To find an ES solution, we consider a homogeneous and isotropic universe described by the Friedmann-Lema$\hat{\imath}$tre-Robertson-Walker metric
\bea
ds^{2}=a(\eta)^2 [- d\eta^{2}+ \gamma_{ij} dx^{i} dx^{j}],
\eea
where $\eta$ is the conformal time, $a(\eta)$ denotes the conformal scale factor, and $\gamma_{ij}$ represents the metric on the three-sphere
\bea
\gamma_{ij} dx^{i} dx^{j}=\frac{dr^{2}}{1-K r^{2}}+r^{2}(d\theta^2+\sin^{2}\theta d\phi^2) .
\eea
Here $K=+1, 0, -1$ corresponds to a closed, flat, and open universe, respectively. The ($00$) and ($ij$) components of  Eq.~(\ref{Einstein}) give
\bea\label{00}
\mathcal{H}^{2}+K=\frac{8\pi G}{3} a^{2} \Big[\rho+\frac{1}{2a^{2}}\phi'^{2}+V-\kappa\frac{3}{2a^{4}}(3\mathcal{H}^{2}+K)\phi'^{2}\Big],
\eea
\bea\label{ij}
2\mathcal{H}'+\mathcal{H}^{2}+K=-8\pi G a^{2}\Big[p+\frac{1}{2 a^{2}}\phi'^{2}-V+\kappa\frac{1}{2 a^{4}}\big(2\mathcal{H}'-3\mathcal{H}^{2}-K+4\mathcal{H}\frac{\phi''}{\phi'}\big)\phi'^{2}\Big],
\eea
where $\rho$ and $p$ are the energy density and the pressure of the perfect fluid, respectively, $p=w\rho$ with $w$ being a constant, $\mathcal{H}=\frac{1}{a}\frac{da}{d\eta}$ and $'$ denotes the derivative with respect to the conformal time $\eta$.  From Eq.~(\ref{scalar}) we obtain the dynamical equation of the scalar field
\bea\label{s}
\frac{1}{a^{2}}(\phi''+2\mathcal{H}\phi')-\kappa\frac{3}{a^{4}}\big[(\mathcal{H}^{2}+K)\phi''+2\mathcal{H}\mathcal{H}'\phi'\big]=-V_{,\phi}.
\eea

\subsection{Einstein static solution}

The ES solution requires that the conditions of $a=a_{0}=constant$ and $a'_{0}=a''_{0}=0$ should be satisfied. Then, Eq.~(\ref{00}) can be reduced to 
\bea\label{11}
\frac{K}{a_{0}^{2}}=\frac{8\pi G}{3} \Big(\rho_{0}+\frac{1}{2a_{0}^{2}}\phi'^{2}_{0}+V_{0}-\frac{3}{2a_{0}^{4}}\kappa K\phi'^{2}_{0}\Big),
\eea
where the subscript $0$ represents the value at the ES state. It is easy to see  that to obtain an ES state $\rho_{0}$, $V_{0}$ and $\phi'^{2}_{0}$ must be  constant.   From Eqs.~(\ref{ij}) and (\ref{s}) we have
\bea\label{12}
\frac{K}{a_{0}^{2}}=-8\pi G \Big(p_{0}+\frac{1}{2a_{0}^{2}}\phi'^{2}_{0}-V_{0}-\frac{1}{2a_{0}^{4}}\kappa K\phi'^{2}_{0}\Big),
\eea
\bea
\frac{dV}{d\phi}|_{\phi=\phi_{0}}=0.
\eea
Thus,  the scalar field with a constant speed moves on a constant potential in the ES state. However, in \cite{Atazadeh2015}, a special case $\phi'_{0}=0$ was considered and $\phi_0=0$ was assumed. Combining Eqs.~(\ref{11}) and (\ref{12}) leads to
\bea\label{es0}
\frac{K}{a^{2}_{0}}=4\pi G \rho_{0}(1+w)+4\pi G \phi'^{2}_{0}\frac{1}{a^{2}_{0}}-8\pi G \phi'^{2}_{0}\kappa\frac{K}{a^{4}_{0}},
\eea
which indicates that $K\neq 0$  for the existence of an ES solution since $\frac{1}{a^{2}} \phi'^{2}=\big(\frac{d\phi}{dt}\big)^{2}$. After introducing two new constants
\bea\label{ll}
F=4\pi G \phi'^{2}_{0} \frac{1}{a^{2}_{0}}=4\pi G \dot{ \phi}^{2}_{0}, 
 \qquad \rho_{s}=4\pi G  \rho_{0}, 
\eea
Eq.~(\ref{es0}) can be re-expressed as
\bea
\frac{1}{a^{2}_{0}}=\frac{F+(1+w)\rho_{s}}{(1+2\kappa F)K}.
\eea
Since $a_{0}$  and $\rho_{0}$ should take positive values, the existence conditions of ES solutions are $a^{2}_{0}>0$ and $\rho_{s}>0$. For the case $\kappa<0$, when $K=1$, we find that  the existence of ES solutions requires
\bea\label{ex1}
 && w<-1, \qquad 0<F<-\frac{1}{2\kappa}, \qquad 0<\rho_{s}<-\frac{F}{1+w},\nonumber\\
 && w<-1, \qquad F>-\frac{1}{2\kappa}, \qquad \rho_{s}>-\frac{F}{1+w},\nonumber\\
 \quad && w=-1, \qquad 0<F<-\frac{1}{2\kappa}, \qquad \rho_{s}>0,\nonumber\\
or \quad && w>-1, \qquad 0\leq F<-\frac{1}{2\kappa}, \qquad \rho_{s}>0.
\eea
While, for $K=-1$, the conditions become
\bea
&& w<-1, \qquad 0\leq F<-\frac{1}{2\kappa}, \qquad \rho_{s}>-\frac{F}{1+w},\nonumber\\
&& w<-1, \qquad F>-\frac{1}{2\kappa}, \qquad 0<\rho_{s}<-\frac{F}{1+w},\nonumber\\
or \quad && w\geq-1, \qquad F>-\frac{1}{2\kappa}, \qquad \rho_{s}>0.
\eea
For the case of $\kappa>0$,  $a^{2}_{0}>0$ and $\rho_{s}>0$ lead to
\bea\label{ex2}
&& w<-1, \qquad F>0, \qquad 0<\rho_{s}<-\frac{F}{1+w},\nonumber\\
 \quad  && w=-1, \qquad F>0, \qquad \rho_{s}>0,\nonumber\\
or \quad  && w>-1, \qquad F\geq 0, \qquad \rho_{s}>0,
\eea
when $K=1$,  and
\bea
w<-1, \qquad F\geq0, \qquad \rho_{s}>-\frac{F}{1+w},
\eea
when $K=-1$.

In the following we will discuss the stability  of ES solutions under the scalar and tensor perturbations. The tensor perturbations will be analyzed firstly since they are relatively easy to handle.

\section{Tensor perturbations}

For the tensor perturbations, the perturbed metric has the following form~\cite{Bardeen1980}
\bea\label{ds0}
ds^{2}=a(\eta)^{2}[-d\eta^{2}+(\gamma_{ij}+2 h_{ij} )dx^{i}dx^{j}].
\eea
For convenience, we perform a harmonic decomposition for the perturbed variable $ h_{ij}$
\bea
 h_{ij}=H_{T, nlm}(\eta) Y_{ij, nlm}(\theta^k),
 \eea
 where summations over $n$, $m$, $l$ are implied. The quantum numbers $m$ and $l$ will be suppressed hereafter as they do not enter the differential equation for the perturbations. The harmonic function $Y_{n}=Y_{nlm}(\theta^{i})$ satisfies~\cite{Harrison1967}
\bea\label{Yn}
\Delta Y_{n}=-k^{2}Y_{n}=\Bigg\{
\begin{array}{rrr}
-n(n+2)Y_{n}, \quad n=0,1,2,..., \quad K=+1\\
-n^{2}Y_{n}, \quad\quad\quad\quad n^{2}\geq 0, \quad\quad\quad\quad K=0\\
-(n^{2}+1)Y_{n}, \quad\quad n^{2}\geq 0, \quad\quad\quad K=-1
\end{array}
\eea
Here, $\Delta$ represents the 3-dimensional spatial Laplacian operator. The spectrum of the perturbation modes is discrete for $K=1$, while it is continuous for $K=0$ or $-1$.   

Substituting the metric given in Eq.~(\ref{ds0}) into the field equations (Eq.~(\ref{Einstein})) leads to \bea
\big(1+4\pi G \phi'^{2}\frac{\kappa}{a^{2}}\big)H''_{T}+\big(2\mathcal{H}+8\pi G \phi'\phi''\frac{\kappa}{a^{2}}\big)H'_{T}+\big(1-4\pi G \phi'^{2}\frac{\kappa}{a^{2}}\big)(k^{2}+2K)H_{T}=0.
\eea
Under the ES background, this equation can be simplified as
\bea
H''_{T}+B H_{T}=0, \qquad B\equiv \frac{(1-\kappa F)}{(1+\kappa F)}(k^{2}+2K).
\eea
To obtain the stable ES solution against tensor perturbations, $B>0$ must be satisfied for any $k$. 
For the case of $\kappa<0$, when $K=1$, we find that the restriction condition $B>0$ gives
\bea\label{t1}
\qquad k^{2}\geq 0, \qquad 0\leq F<-\frac{1}{\kappa}.
\eea
While, when $K=-1$, the stable ES solution requires
\bea\label{t11}
  \qquad k^{2}>2, \qquad 0\leq F<-\frac{1}{\kappa},\nonumber\\
or  \qquad 1\leq k^{2}<2, \qquad F>-\frac{1}{\kappa}.
\eea

For the case of $\kappa>0$,   $B>0$ leads to
\bea\label{t2}
 k^{2} \geq 0, \qquad 0\leq F<\frac{1}{\kappa},
\eea
when $K=1$, and
\bea\label{t22}
  k^{2}>2, \qquad 0\leq F<\frac{1}{\kappa},\nonumber\\
or \qquad   1\leq k^{2}<2, \qquad F>\frac{1}{\kappa}.
\eea
when $K=-1$.
It is easy to see that the stable ES solution exists only  in the case of spatially closed universe ($K=1$). Thus, in the following analysis $K=1$ is considered. Combing the existence conditions given in Eqs.~(\ref{ex1},\ref{ex2}) and the stability conditions under tensor perturbations, we obtain that $w$, $F$ and $\rho_s$ should satisfy
\bea\label{ext1}
 && w<-1, \qquad 0<F<-\frac{1}{2\kappa}, \qquad 0<\rho_{s}<-\frac{F}{1+w},\nonumber\\
 && w<-1, \qquad -\frac{1}{2\kappa}<F<-\frac{1}{\kappa}, \qquad \rho_{s}>-\frac{F}{1+w},\nonumber\\
 && w=-1, \qquad 0<F<-\frac{1}{2\kappa}, \qquad \rho_{s}>0,\nonumber\\
or \quad && w>-1, \qquad 0\leq F<-\frac{1}{2\kappa}, \qquad \rho_{s}>0,
\eea
for $\kappa<0$,  and
\bea\label{ext2}
&& w<-1, \qquad 0<F<\frac{1}{\kappa}, \qquad 0<\rho_{s}<-\frac{F}{1+w},\nonumber\\
&& w=-1, \qquad 0<F<\frac{1}{\kappa}, \qquad \rho_{s}>0,\nonumber\\
or \quad  && w>-1, \qquad 0\leq F<\frac{1}{\kappa}, \qquad \rho_{s}>0,
\eea
for $\kappa>0$.

\section{Scalar perturbations}

To analyze the stability of ES solutions under scalar perturbations, we consider the  perturbed metric: 
\bea\label{ds}
ds^{2}=a(\eta)^{2} [-(1+2\Psi)d\eta^{2}+ (1+2\Phi) \gamma_{ij} dx^{i} dx^{j}],
\eea
where the Newton gauge has been used, $\Psi$ is the Bardeen potential and $\Phi$ denotes the perturbation to the spatial curvature.

Using the above perturbed metric and the field equations given in Eqs.~(\ref{Einstein}, \ref{scalar}), we obtain the following perturbation equations
\bea\label{p1}
&&\frac{1}{4\pi G a^{2}_{0}}(\nabla^{2}\Phi+3 \Phi)=-\delta\rho+\frac{1}{a^{2}_{0}}(\phi'^{2}_{0} \Psi-\phi'_{0} \delta\phi')\nonumber\\
&&\qquad\qquad\qquad\qquad\qquad+\kappa\frac{1}{a^{4}_{0}}[3 \phi'_{0} \delta\phi'-\phi'^{2}_{0}\nabla^{2}\Phi-3 (\Psi+\Phi)\phi'^{2}_{0}],
\eea
\bea\label{p2}
-(\Psi+\Phi)=4\pi G \phi'^{2}_{0} \kappa\frac{1}{a^{2}_{0}} (\Psi-\Phi),
\eea
\bea\label{p3}
&&\frac{3}{a^{2}_{0}}(-\Phi''+ \Phi)+\frac{1}{a^{2}_{0}}\nabla^{2}(\Psi+\Phi)=4\pi G \bigg\{3\delta p-\frac{3}{a^{2}_{0}}(\phi'^{2}_{0}\Psi-\phi'_{0} \delta\phi')\nonumber\\
&&\qquad\qquad+\kappa\frac{1}{a^{4}_{0}}[\phi'^{2}_{0}\nabla^{2}(\Phi-\Psi)+3\phi'^{2}_{0}\Phi''+3(\Psi+\Phi)\phi'^{2}_{0}-3\phi'_{0} \delta\phi']\bigg\}
\eea
\bea\label{p4}
\bigg(1-\kappa\frac{3}{a^{2}_{0}}\bigg)\delta\phi''-\bigg(1-\kappa\frac{1}{a^{2}_{0}}\bigg)\nabla^{2}\delta\phi-
\bigg[(\Psi'-3\Phi')-\kappa\frac{3}{a^{2}_{0}}(\Psi'-\Phi')\bigg]\phi'_{0}=0.
\eea
Here, the perturbation of the scalar field $\phi\rightarrow \phi_{0}+\delta \phi$ is considered. For the perfect fluid, the perturbation of its energy-momentum tensor can be expressed as
\bea
\delta T^{\mu(m)}_{\nu}=\delta\rho u^{\mu}u^{\nu}+u^{\mu}D_{\nu}q+u_{\nu}D^{\mu}q+\delta p P^{\mu}_{\nu},
\eea
where $u^{\mu}$ is the four-velocity of matter and $q$ is related to the perturbation of the spatial component of this four-velocity. The projection tensor $P^{\mu}_{\nu}$ and the derivative $D_{\mu}$ are defined as
\bea
&&P^{\mu}_{\nu}=\delta^{\mu}_{\nu}+u^{\mu}_{\nu},\\
&&D_{\mu}=P^{\alpha}_{\mu}\partial_{\alpha}=\partial_{\mu}+u_{\mu}u^{\alpha}\partial_{\alpha}.
\eea
The relation between the density and pressure perturbations is
\bea\label{rhop}
\delta p=c^{2}_{s} \rho_{0}\delta,
\eea
where, $\delta=\delta\rho/\rho_{0}$ and $c^{2}_{s}=w$ is the sound speed.

Similar to the case of  tensor perturbations, we perform a harmonic decomposition for the perturbed variables
\bea\label{Y}
&&\Psi=\Psi_{n}(\eta)Y_{n}(\theta^{i}), \qquad \Phi=\Phi_{n}(\eta)Y_{n}(\theta^{i}), \qquad q=q_{n}(\eta)Y_{n}(\theta^{i}),\nonumber\\
&&\delta=\delta_{n}(\eta)Y_{n}(\theta^{i}), \qquad \delta\phi=\delta\phi_{n}(\eta)Y_{n}(\theta^{i}).
\eea

Combining Eqs~(\ref{p1}),~(\ref{p2}),~(\ref{p3}) and~(\ref{p4}) gives two independent perturbed equations
\bea\label{ie1}
\Phi''_{n}+b_{11}\Phi_{n}+a_{12}\delta\phi'_{n}=0,
\eea
\bea\label{ie2}
\delta\phi''_{n}+b_{22}\delta\phi_{n}+a_{21}\Phi'_{n}=0,
\eea
with
\bea\label{ba}
&&b_{11}=w k^{2}-\frac{\kappa F(1-\kappa F)(w-1)}{\frac{\kappa}{a^{2}_{0}}(1+\kappa F)^{2}}-\frac{2\kappa^{2} F^{2}(3w-1)+(1+\kappa F)(3w+1)}{(1+\kappa F)^{2}},\nonumber\\
&&a_{12}=-\frac{F[(1-3w)\frac{\kappa}{a^{2}_{0}}+(w-1)]}{\frac{1}{a^{2}_{0}}(1+\kappa F)}\frac{1}{\phi'_{0}},\nonumber\\
&&b_{22}=\frac{1-\frac{\kappa}{a^{2}_{0}}}{1-3\frac{\kappa}{a^{2}_{0}}}k^{2},\nonumber\\
&&a_{21}=2\Big(\frac{1}{1-3\frac{\kappa}{a^{2}_{0}}}+\frac{1}{1+\kappa F}\Big)\phi'_{0}.
\eea

Introducing two new variables $\delta\varphi=\delta\phi'$ and $\Upsilon=\Phi'$,  Eqs.~(\ref{ie1}) and~(\ref{ie2}) can be rewritten as
\bea
&&\Phi'_{n}-\Upsilon_{n}=0,\nonumber\\
&&\Upsilon'_{n}+b_{11}\Phi_{n}+a_{12}\delta\varphi_{n}=0,\nonumber\\
&&\delta\phi'_{n}-\delta\varphi_{n}=0,\nonumber\\
&&\delta\varphi'_{n}+b_{22}\delta\phi_{n}+a_{21}\Upsilon_{n}=0.
\eea
The stability of ES solutions is determined by the eigenvalues of the coefficient matrix, which is
\bea
\mu^{2}=\frac{-M\pm\sqrt{N}}{2},
\eea
where
\bea
M=b_{11}+b_{22}-a_{12}a_{21}, \qquad N=-4 b_{11}b_{22}+(b_{11}+b_{22}-a_{12}a_{21})^{2}.
\eea
If $\mu^{2}<0$, a small perturbation from the ES state will result in an oscillation around this state  rather than an exponential deviation. Thus, the corresponding ES solution is stable. Otherwise, it is unstable. $\mu^{2}<0$ gives the stability conditions under   scalar perturbations
\bea\label{MN}
M>0, \qquad N>0, \qquad M^{2}>N.
\eea
Since  $b_{22}=0$ and $M^{2}=N$ when $k^{2}=0$, the homogeneous scalar perturbations require
\bea\label{M}
\mu^{2}=\frac{-M-\sqrt{N}}{2}=-M=-b_{11}+a_{12}a_{21}<0.
\eea

\subsection{Stability}

For the scalar perturbations, the analysis of  the stability of ES solutions is very complicated. To simplify discussions, we will consider  the constraints from the tensor perturbations  and the existence conditions obtained in the previous sections, in which it is found that  the ES solution is stable under 
the conditions of $K=1$ and Eq.~(\ref{ext1}, \ref{ext2}). 

\subsubsection{$\kappa<0$}

From Eq.~(\ref{M}), we obtain that  the stability conditions   under  homogeneous scalar perturbations are
\bea
&& F=0,\qquad -1<w<-\frac{1}{3}, \qquad \rho_{s}>0,\nonumber\\
or \quad && 0<F<-\frac{1}{4\kappa}, \qquad -1<w<\frac{-1+2\kappa F}{3+6\kappa F}, \qquad \rho_{s}>\lambda_{+},
\eea
where
\bea
&&\lambda_{\pm}=\frac{1+3w+\kappa F[-11+3w-2\kappa F(1+3w)]}{6\kappa (1+w)[1+3w+\kappa F(-2+6w)]}\nonumber\\
&&\qquad\pm \frac{1}{6}\sqrt{\frac{(1+2\kappa F)^{2}[1+(22-23\kappa F)\kappa F+6w+6\kappa F(10+13\kappa F)w+9(-1+\kappa F)^{2} w^{2}]}{\kappa^{2}(1+w)^{2}[1+3w+\kappa F(-2+6w)]^{2}}}.\nonumber\\
\eea
For $0<F<-\frac{1}{4\kappa}$, one can obtain $-1<\frac{-1+2\kappa F}{3+6\kappa F}<-\frac{1}{3}$, which means that  $w$ is negative.

For the inhomogeneous scalar perturbations,  the physical modes have $n\geq 2$ which gives $k^{2}\geq 8$ since the  $n=1$ mode corresponds to a gauge degree of freedom related to a global rotation.
For $F=0$ and $k^{2}=8$, we obtain that the region of $w$ and $\rho_{s}$
\bea
&&\frac{1}{5}<w\leq\frac{11}{15}, \quad \rho_{s}>0,\nonumber\\
&& \frac{11}{15}<w<\frac{9}{5}, \quad 0<\rho_{s}<\frac{-9+5w}{-11\kappa+4\kappa w+15\kappa w^{2}},\nonumber\\
&& \frac{11}{15}<w<\frac{9}{5}, \quad \rho_{s}>\frac{-9+5w}{-11\kappa+4\kappa w+15\kappa w^{2}},\nonumber\\
or \quad && w\geq\frac{9}{5}, \quad \rho_{s}>0,
\eea
While, when $0<F<-\frac{1}{4\kappa}$ and $k^{2}=8$,  we find that $w$ and $\rho_{s}$ need to satisfy
\bea
&& 0<w<\frac{1+\kappa F-2\kappa^{2}F^{2}}{5+13\kappa F+2\kappa^{2}F^{2}}, \nonumber\\
\quad && 0<\rho_{s}<\frac{-4F w[1+3\kappa F+\kappa^{2}F^{2}]}{(1+w)[-1+5w+\kappa F(-1+13w)+2\kappa^{2}F^{2}(1+w)]},\nonumber\\
or \quad && w\geq \frac{1+\kappa F-2\kappa^{2}F^{2}}{5+13\kappa F+2\kappa^{2}F^{2}}>0, \quad \rho_{s}>0.
\eea
Obviously,  a positive $w$ is required for the stable ES solution under  the inhomogeneous scalar perturbations, which conflicts with the conditions given by the homogeneous scalar perturbations and the tensor ones. Thus, there is no stable ES solution in the case of $\kappa<0$.

\subsubsection{$\kappa>0$}

When $\kappa>0$, the results are   summarized in Tab.~(\ref{Tab1}) where  the conditions shown in Eq.~(\ref{ext2}) have been considered together. The constants $\xi$ and $\zeta$ are defined as
\bea
&&\xi=\frac{-11-30w+9w^{2}}{\kappa(-23+78w+9w^{2})}+12\sqrt{\frac{1+5w+3w^{2}-9w^{3}}{\kappa^{2}(-23+78w+9w^{2})^{2}}},\nonumber\\
&&\zeta=\frac{4F(1+\kappa F+\kappa^{2}F^{2})}{(1+w)[-1-3w+\kappa F(11-3w+2\kappa F+6\kappa F w)]}.\nonumber
\eea

\begin{table}
\caption{\label{Tab1} Summary of the combinations of the stability conditions under homogeneous scalar perturbations and that given in Eq.~(\ref{ext2}) with $K=1$ and $\kappa>0$.}
\begin{center}
 \begin{tabular}{|r|r|r|}
  \hline
  \hline
  $w$ & $F$ & $\rho_{s}$\\
  \hline
  $-1< w<-\frac{1}{3}$ & $F=0$ & $\rho_{s}>0$\\
  & $0<F<\xi$ & $\lambda_{-}<\rho_{s}<\lambda_{+}$\\
  & $0<F<\frac{1}{\kappa}$ & $\rho_{s}>\frac{1-\kappa F}{3\kappa+3\kappa w}$\\
  \hline
  $w=-\frac{1}{3}$ & $\quad 0<F<\frac{1}{\kappa}$ & $\quad \rho_{s}>\frac{1-\kappa F}{2\kappa}$\\
  \hline
  $-\frac{1}{3}<w<0$ & $0<F<\frac{1+3w}{2\kappa-6\kappa w}$ & $\frac{1-\kappa F}{3\kappa+3\kappa w}<\rho_{s}<\lambda_{+}$\\
  & $F=\frac{1+3\kappa w}{2\kappa-6\kappa w}$ & $\frac{1-\kappa F}{3\kappa+3\kappa w}<\rho_{s}<\zeta$\\
  & $\frac{1+3\kappa w}{2\kappa-6\kappa w}<F\leq\xi$ & $\frac{1-\kappa F}{3\kappa +3\kappa w}<\rho_{s}<\lambda_{-}$\\
  & $\frac{1+3w}{2\kappa-6\kappa w}<F<\xi$ & $\rho_{s}>\lambda_{+}$\\
  & $F=\xi$ & $\rho_{s}>\lambda_{-}$\\
  & $\xi<F<\frac{1}{\kappa}$ & $\rho_{s}>\frac{1-\kappa F}{3\kappa+3\kappa w}$\\
  \hline
  $0\leq w<\frac{1}{9}$ & $0<F<\frac{1+3w}{2\kappa-6\kappa w}$ & $\frac{1-\kappa F}{3\kappa+3\kappa w}<\rho_{s}<\lambda_{+}$\\
  & $F=\frac{1+3w}{2\kappa-6\kappa w}$ & $\frac{1-\kappa F}{3\kappa+3\kappa w}<\rho_{s}<\zeta$\\
  & $\frac{1+3w}{2\kappa-6\kappa w}<F<\frac{1}{\kappa}$ & $\frac{1-\kappa F}{3\kappa+3\kappa w}<\rho_{s}<\lambda_{-}$\\
  & $\frac{1+3w}{2\kappa-6\kappa w}<F<\frac{1}{\kappa}$ & $\rho_{s}>\lambda_{+}$\\
  \hline
  $w\geq\frac{1}{9}$ & $0<F<\frac{1}{\kappa}$ & $\frac{1-\kappa F}{3\kappa+3\kappa w}<\rho_{s}<\lambda_{+}$\\
 \hline
 \hline
  \end{tabular}
\end{center}
   \end{table}

Now we consider the contribution from inhomogeneous scalar perturbations. We find that there is no stable ES solution for $w\leq 0$ since $M>0$ and $M^{2}-N>0$ can not be satisfied simultaneously  when $k^{2}=8$. Since the expressions are  complicated, we do not show them here.

When $0<w<\frac{1}{9}$, from Tab.~(\ref{Tab1}) one can see that there are four different kinds of stability conditions under  homogeneous scalar perturbations. We will analyze inhomogeneous scalar perturbations under these conditions, respectively.

(i) $0<F<\frac{1+3w}{2\kappa-6\kappa w}$ and $\frac{1-\kappa F}{3\kappa+3\kappa w}<\rho_{s}<\lambda_{+}$. The stability condition $M>0$  under inhomogeneous scalar perturbations requires  $\rho_{s}$ to satisfy
\bea\label{rhow}
0<\rho_{s}<\frac{1-\kappa F}{3\kappa+3\kappa w}, \quad  or \qquad \rho_{s}>\varpi,
\eea
where
\bea
&&\varpi=\frac{7+[19+\kappa F(2-26w)-27w]\kappa F+5w}{2\kappa(1+w)[5+15w+2\kappa F(7+3w)]}\nonumber\\
&&\quad+\frac{1}{2}\sqrt{\frac{(1+2\kappa F)^{2}[49+3\kappa F(50+19\kappa F)+70w-2\kappa F(34+57\kappa F)w+(5+11\kappa F)^{2} w^{2}]}{\kappa^{2}(1+w)^{2}[5+15w+2\kappa F(7+3w)]^{2}}},\nonumber\\
\eea
where $k^2=8$ is taken. Since $\varpi>\lambda_{+}$  for $0<w<\frac{1}{9}$ and $0<F<\frac{1+3w}{2\kappa-6\kappa w}$, there is no overlap for the allowed regions of $\rho_{s}$ from homogeneous and inhomogeneous scalar perturbations, which indicates that the ES solution is unstable.

(ii)  $F=\frac{1+3w}{2\kappa-6\kappa w}$ and $\frac{1-\kappa F}{3\kappa+3\kappa w}<\rho_{s}<\zeta$. The inhomogeneous perturbations   require $\rho_{s}$ to satisfy Eq.~(\ref{rhow}) when $k^2=8$. Since $\varpi>\zeta$,  there is no stable ES solution in this case too.

\begin{figure}[!htb]
                \centering
                \includegraphics[width=0.457\textwidth ]{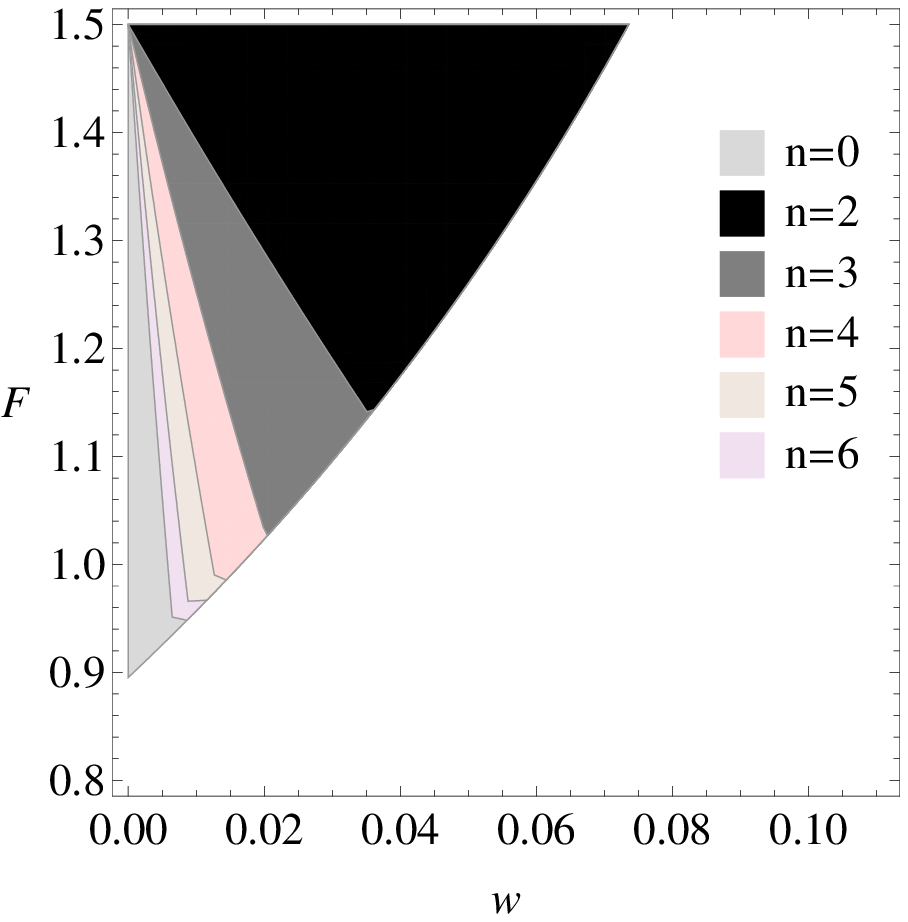}
                \includegraphics[width=0.457\textwidth ]{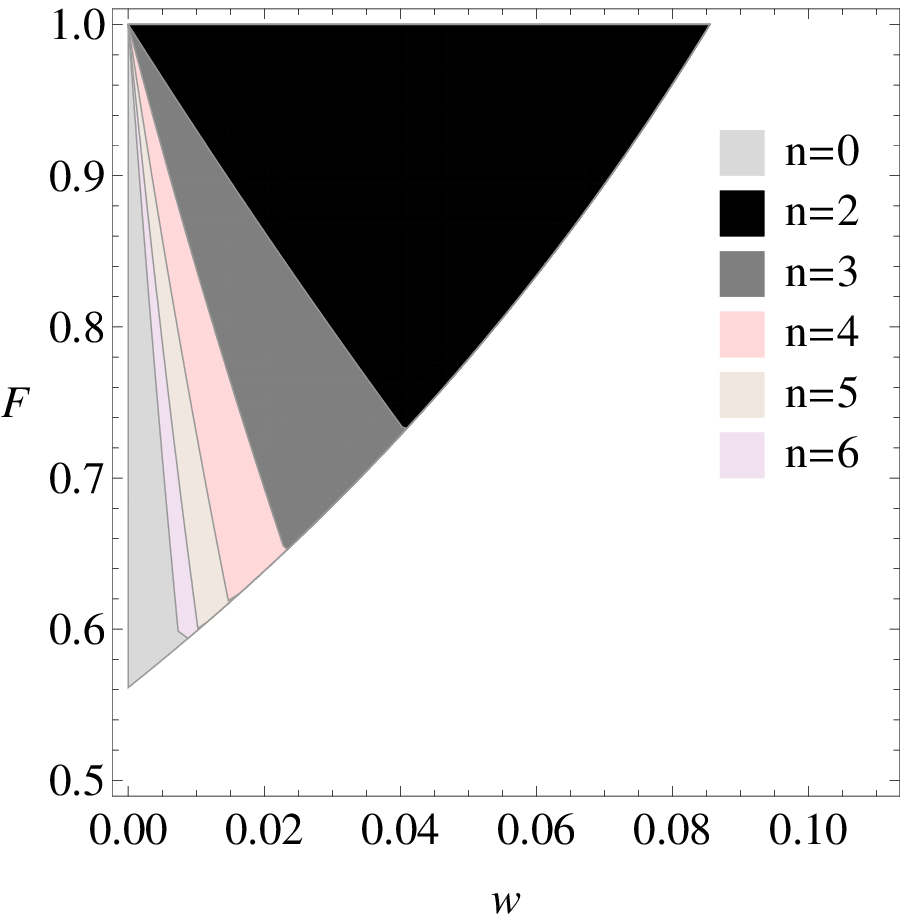}
                \caption{\label{Fig1} The stability regions in the  $F-w$ plane under homogeneous   and inhomogeneous perturbations with $n$ taken to be  $n=0, 2,3,4,5,6$. $n=0$ corresponds to the results from homogeneous perturbations.  The left panel is plotted with $\rho_{s}=15$ and $\kappa=\frac{2}{3}$, while the right one for $\rho_{s}=15$ and $\kappa=1$. }
        \end{figure}

(iii) $\frac{1+3w}{2\kappa-6\kappa w}<F<\frac{1}{\kappa}$ and $\frac{1-\kappa F}{3\kappa+3\kappa w}<\rho_{s}<\lambda_{-}$. When $k^2=8$, $\rho_s$ is also required to satisfy Eq.~(\ref{rhow}). We find that  $\varpi>\lambda_{-}$. Thus, the ES solution is unstable.

(iv) $\frac{1+3w}{2\kappa-6\kappa w}<F<\frac{1}{\kappa}$ and $\rho_{s}>\lambda_{+}$. In this case,  since the analytical results for the stability regions under inhomogeneous scalar perturbations  are very complicated, we do not show them here  and  but resort to a numerical discussion. We find that when $n=2$ the smallest stability regions is obtained. With the  increase of  the value of $n$,  the stability regions become larger and larger, which can be seen from Fig.~(\ref{Fig1}).  In this Figure,  $n$ is taken to be $n=0,2,3,4,5,6$, respectively, where $n=0$ corresponds to the case of homogeneous scalar perturbations. When $n\rightarrow \infty$,  $b_{11}$ reduces to   $b_{11}\simeq w k^{2}$.  The stable ES solution requires
\bea
M\simeq \Big(w+\frac{1-\frac{\kappa}{a^{2}_{0}}}{1-3\frac{\kappa}{a^{2}_{0}}}\Big)k^{2}>0,\\
M^{2}-N\simeq \frac{4w(1-\frac{\kappa}{a^{2}_{0}})}{1-3\frac{\kappa}{a^{2}_{0}}}k^{4}>0.
\eea
The above two equations give
\bea
-\frac{F}{1+w}<\rho_{s}<\frac{1-\kappa F}{3\kappa+3\kappa w}, \qquad \rho_{s}>\frac{1+\kappa F}{\kappa+\kappa w}.
\eea
where $0<w<\frac{1}{9}$ and $\frac{1+3w}{2\kappa-6\kappa w}<F<\frac{1}{\kappa}$ are considered together. Since $\frac{1+\kappa F}{\kappa+\kappa w}<\lambda_{+}$ is always satisfied, the stability regions are larger than what are obtained under homogeneous  scalar perturbations.
Thus, in this case  the ES solution is stable under both  homogeneous and inhomogeneous scalar perturbations and the stability regions are given by taking $n=2$.

When $w\geq \frac{1}{9}$, we find that  $M>0$ and $M^{2}-N>0$ can not be satisfied simultaneously, which shows that the stable  ES solution does not exist.

\section{Conclusion}

In this paper, we have analyzed the stability of an ES universe under  both scalar and tensor perturbations in gravity theory  with  a  coupling between the kinetic term of the scalar field and the  Einstein tensor.  Homogeneous and inhomogeneous perturbations are considered together and  inhomogeneous perturbations  will compress the allowed regions of model parameters significantly. We find that  the stable ES solution exists only  in the spatially  closed universe ($K=1$) and it requires  the coupling constant   $\kappa>0$ to be positive. In addition, the equation of state of the perfect fluid is required to satisfy $0<w<\frac{1}{9}$, which indicates that if this perfect fluid is the pressureless matter or radiation the stable ES solution does not exist, although it does under homogeneous perturbations. Thus,  in the non-minimally kinetic coupled gravity with the perfect fluid satisfying  $0<w<\frac{1}{9}$ the stable ES solution can exist under both scalar and tensor perturbations  and the emergent mechanism can be used to avoid the big bang singularity.

When $F=0$, our results reduce to what were obtained in \cite{Atazadeh2015} where a special condition  $\dot{\phi}=0$ was considered. In this special case   our analyses show that inhomogeneous scalar perturbations will break the stability of an ES solution although the solution is stable under tensor and homogeneous scalar  ones. Therefore, the big bang singularity problem can not be solved successfully if $\dot{\phi}=0$ is taken.

Finally, a few comments are now in order for the emergent scenario proposed  for avoiding the big-bang singularity.  The emergent scenario assumes  the existence of a stable Einstein static state and its past eternity.  But usually such a state only exists under certain conditions. So, there is a question as to how this particular state comes into being in the first place,  and in this regard, let us note that  one possibility might be  the creation of this state  from ``nothing" through quantum tunneling\cite{Hartle:1983ai,Vilenkin:1984wp}.  Another issue is that even 
this state is stable classically, one still needs to address the question as to whether it is stable quantum mechanically, possibly by  calculating the  characteristic decay time in a quantum theory of cosmology when  this state was formed.

\begin{acknowledgments}
This work was supported by the National Natural Science Foundation of China under Grants No. 11775077,  No. 11435006, No.11690034, and  No. 11375092.

\end{acknowledgments}


\begin{thebibliography}{00}


\bibitem{Staro1980}A. A.  Starobinsky, Phys. Lett. B {\bf 91}, 99 (1980).
\bibitem{Guth1981}A. H. Guth, Phys. Rev. D {\bf 23}, 347 (1981).
\bibitem{Linde1982}A. D. Linde, Phys. Lett. B {\bf 108},  389 (1982).

\bibitem{Gasperini2003}
M. Gasperini and G. Veneziano, Phys. Rep. {\bf 373}, 1 (2003);
J. E. Lidsey, D. Wands, and E. J. Copeland, Phys. Rep. {\bf 337}, 343 (2000).

\bibitem{Khoury2001}
J. Khoury, B. A. Ovrut, P. J. Steinhardt, and N. Turok, Phys. Rev. D {\bf 64}, 123522 (2001);
P. J. Steinhardt and N. Turok, Science {\bf 296}, 1436 (2002); Phys. Rev. D {\bf 65}, 126003 (2002);
J. Khoury, P. J. Steinhardt, and N. Turok, Phys. Rev. Lett. {\bf 92}, 031302 (2004).

\bibitem{Ellis2004}
G. F. R. Ellis and R. Maartens, Class. Quant. Grav. {\bf 21}, 223 (2004);
G. F. R. Ellis, J. Murugan, and C. G. Tsagas,  Class. Quant. Grav.  {\bf 21}, 233 (2004).

\bibitem{Barrow2003}
J. D. Barrow, G. F. R. Ellis, R. Maartens and C. G. Tsagas, Class. Quant. Grav. {\bf 20}, L155 (2003);
A. S. Eddington, Mon. Not. Roy. Astron. Soc. {\bf 90}, 668 (1930);
G. W. Gibbons, Nucl. Phys. B {\bf 292}, 784 (1987);
G. W. Gibbons, Nucl. Phys. B {\bf 310}, 636 (1988).

\bibitem{Abbott}   L. F. Abbott,  Nucl. Phys. B {\bf 185}, 233 (1981); T. Futamase and K. i. Maeda, Phys. Rev. D {\bf 39}, 399 (1989).

\bibitem{Sahni} V. Sahni and S. Habib, Phys. Rev. Lett. {\bf 81},  1766   (1998);
J. P. Uzan, Phys. Rev. D {\bf 59},  123510 (1999); F. Perrotta, C. Baccigalupi  and S. Matarrese, Phys. Rev. D {\bf 61}, 023507  (1999).
\bibitem{Bergmann} P.G. Bergmann, Int. J. Theor. Phys. {\bf1}, 25 (1968);
                   K. Nordtvedt, Astrophys. J. {\bf161}, 1059 (1970);
                   R. Wagoner, Phys. Rev. D {\bf1}, 3209 (1970).


\bibitem{Staro}  T. P. Sotiriou and V. Faraoni, Rev. Mod. Phys. {\bf 82}, 451 (2010);   A. De Felice and S. Tsujikawa, Living Rev. Relativity {\bf 13}, 3 (2010).
  
  \bibitem{Brans} C. Brans and R. H. Dicke, Phys. Rev. {\bf124}, 925 (1961);
                R. H. Dicke, Phys. Rev. {\bf125}, 2163 (1962).

\bibitem{Planck}Planck Collaboration: P. A. R. Ade, N. Aghanim, M. Arnaud, et al., Astron. Astrophys. {\bf 594}, A20 (2016).

\bibitem{Barrow1983}
J. D. Barrow and A. C. Ottewill, J. Phys. A {\bf 16}, 2757 (1983);
C. G. Bohmer, L. Hollenstein, and F. S. N. Lobo, Phys. Rev. D {\bf 76}, 084005 (2007);
R. Goswami, N. Goheer, and P. K. S. Dunsby, Phys. Rev. D {\bf 78}, 044011 (2008);
N. Goheer, R. Goswami, and P. K. S. Dunsby, Class. Quantum Grav. {\bf 26}, 105003 (2009);
S. del Campo, R. Herrera, and P. Labrana, J. Cosmol. Astropart. Phys. {\bf 07}, 006 (2009).

\bibitem{Seahra2009}
S. S. Seahra and C. G. Bohmer, Phys. Rev. D {\bf 79}, 064009 (2009).

\bibitem{Miao2016}
H. Miao, P. Wu, and H. Yu, Class. Quantum Grav. {\bf 33}, 215011 (2016).

\bibitem{Huang2014} H. Huang, P. Wu, and H. Yu, Phys. Rev. D {\bf 89}, 103521 (2014);
 S. Campo, R. Herrera, and P. Labra\~na, J. Cosmol. Astropart. Phys. {\bf11}, 030 (2007);
                S. Campo, R. Herrera, and P. Labra\~na, J. Cosmol. Astropart. Phys. {\bf07}, 006 (2009).

\bibitem{Li2017} S. Li and H. Wei,  Phys. Rev. D {\bf 96}, 023531 (2017). 

\bibitem{Wu2011}
P. Wu and H. Yu, Phys. Lett. B {\bf 703}, 223 (2011);
J. T. Li, C. C. Lee, and C. Q. Geng, Eur. Phys. J. C {\bf 73}, 2315 (2013).

\bibitem{Mulryne2005}
D. J. Mulryne, R. Tavakol, J. E. Lidsey, and G. F. R. Ellis, Phys. Rev. D {\bf 71}, 123512 (2005);
J. E. Lidsey, D. J. Mulryne, N. J. Nunes, and R. Tavakol, Phys. Rev. D {\bf 70}, 063521 (2004);
L. Parisi, M. Bruni, R. Maartens, and K. Vandersloot, Class. Quantum Grav. {\bf 24}, 6243 (2007);
R. Canonico and L. Parisi, Phys. Rev. D {\bf 82}, 064005 (2010);
P. Wu, S. Zhang, and H. Yu, J. Cosmol. Astropart. Phys. {\bf 05}, 007 (2009);
S. Bag, V. Sahni, Y. Shtanov, and S. Unnikrishnan, J. Cosmol. Astropart. Phys. {\bf 07}, 034 (2014);
M. Khodadi, K. Nozari and E. N. Saridakis, Class. Quant. Grav. {\bf35}, 015010 (2018). 

\bibitem{Zhang2016}
K. Zhang, P. Wu, H. Yu, and L. Luo, Phys. Lett. B {\bf 758}, 37 (2016).

\bibitem{Zhang2010}
K. Zhang, P. Wu, and H. Yu, Phys. Lett. B {\bf 690}, 229 (2010);
K. Zhang, P. Wu, and H. Yu, Phys. Rev. D {\bf 85}, 043521 (2012);
J. E. Lidsey and D. J. Mulryne, Phys. Rev. D {\bf 73}, 083508 (2006);
A. Gruppuso, E. Roessl, and M. Shaposhnikov, J. High Energy Phys. {\bf 08}, 011 (2004);
L. A. Gergely and R. Maartens, Class. Quantum Grav. {\bf 19}, 213 (2002);
K. Atazadeh, Y. Heydarzade, and F. Darabi, Phys. Lett. B {\bf 732}, 223 (2014);
K. Zhang, P. Wu, and H. Yu, J. Cosmol. Astropart. Phys. {\bf 01}, 048 (2014);
Y. Heydarzade, F. Darabi, and K. Atazadeh, Astrophys. Space. Sci. {\bf 361}, 250 (2016);
Y. Heydarzade and F. Darabi, J. Cosmol. Astropart. Phys. {\bf 04}, 028 (2015).

\bibitem{HuangH2015}
H. Huang, P. Wu, and H. Yu, Phys. Rev. D {\bf 91}, 023507 (2015).

\bibitem{Bohmer2009}
C. G. Bohmer and F. S. N. Lobo, Phys. Rev. D {\bf 79}, 067504 (2009).

\bibitem{HuangQ2015}
Q. Huang, P. Wu, and H. Yu. Phys. Rev. D {\bf 91}, 103502 (2015);
C. G. Bohmer, Class. Quantum Grav. {\bf 21}, 1119 (2004);
K. Atazadeh. J. Cosmol. Astropart. Phys. {\bf 06}, 020 (2014).

\bibitem{Wu2010}
P. Wu. and H. Yu, Phys. Rev. D {\bf 81}, 103522 (2010);
C. G. Bohmer and F. S. N. Lobo, Eur. Phys. J. C {\bf 70}, 1111 (2010);
M. Khodadi, Y. Heydarzade, F. Darabi, and E. N. Saridakis. Phys. Rev. D {\bf 93}, 124019 (2016).

\bibitem{Bohmer2015}
C. G. Bohmer, N. Tamanini, and M. Wright, Phys. Rev. D {\bf 92}, 124067 (2015).


\bibitem{Atazadeh2015}
K. Atazadeh and F. Darabi, Phys. Lett. B {\bf 744}, 363 (2015).

\bibitem{HuangH2014}
H. Huang, P. Wu, and H. Yu, Phys. Rev. D {\bf 89}, 103521 (2014).

\bibitem{Bohmer2013}
C. G. Bohmer, F. S. N. Lobo, and N. Tamanini, Phys. Rev. D {\bf 88}, 104019 (2013).



\bibitem{Tawfik2016}
A. N. Tawfik, A. M. Diab, E. A. E. Dahab, and T. Harko, Phys. Rev. D {\bf 93}, 063526 (2016);
 arXiv:1608.06532;
M. Khodadi, Y. Heydarzade, K. Nozari, and F. Darabi, Eur. Phys. J. C {\bf 75}, 590 (2015);
S. Carneiro and R. Tavakol, Phys. Rev. D {\bf 80}, 043528 (2009);
A. Odrzywolek, Phys. Rev. D {\bf 80}, 103515 (2009);
T. Clifton and J. D. Barrow, Phys. Rev. D {\bf 72}, 123003 (2005);
A. Vilenkin, Phys. Rev. D {\bf 88}, 043516 (2013);
A. Aguirre and J. Kehayias, Phys. Rev. D {\bf 88}, 103504 (2013);
A. T. Mithani and A. Vilenkin, J. Cosmol. Astropart. Phys. {\bf 01}, 028 (2012);
Y. Cai, Y. Wan, and X. Zhang, Phys. Lett. B {\bf 731}, 217 (2014);
Y. Cai, M. Li and X. Zhang, Phys. Lett. B {\bf 718}, 248 (2012);
M. Mousavi and F. Darabi, Nucl. Phys. B {\bf 919}, 523 (2017);
H. Shabani and A. H. Ziaie, Eur. Phys. J. C {\bf 77}, 31 (2017);
F. Darabi, K. Atazadeh, arXiv:1704.03040.

\bibitem{Amendola1993}
L. Amendola, Phys. Lett. B {\bf 301}, 175 (1993).

\bibitem{Capozziello1999}
S. Capozziello and G. Lambiase, Gen. Rel. Grav. {\bf 31}, 1005 (1999);
S. Capozziello, G. Lambiase, and H. J. Schmidt, Annalen Phys. {\bf 9}, 39 (2000).

\bibitem{Granda2011}
L. N. Granda, JCAP {\bf 04}, 016 (2011).

\bibitem{Yang2016}
N. Yang, Q. Fei, Q. Gao, and Y. Gong, Class. Quantum Grav. {\bf 33}, 205001 (2016);
Y. Huang, Q. Gao, and Y. Gong,  	Eur. Phys. J. C {\bf 75}  143 (2015);
C. Germani and A. Kehagias, JCAP {\bf 05}, 019 (2010);
C. Germani and Y. Watanabe, JCAP {\bf 07}, 031 (2011);
S. Tsujikawa, Phys. Rev.D {\bf 85}, 083518 (2012);
H. M. Sadjadi and P. Goodarzi, Phys. Lett. B {\bf 732}, 278 (2014).

\bibitem{Darabi2015}
F. Darabi, and A. Parsiya, Class. Quantum Grav. {\bf 32}, 155005 (2015).
\bibitem{Sush2009}S V. Sushkov,   Phys. Rev. D {\bf 80}, 103505 (2009).
\bibitem{Gao2010}
C. Gao, JCAP {\bf 06}, 023 (2010);
A. Ghalee, Phys. Rev. D {\bf 88}, 083528 (2013).

\bibitem{Granda2010}
L. N. Granda, JCAP {\bf 07}, 006 (2010);
L. N. Granda and W. Cardona, JCAP {\bf 07}, 021 (2010);
L. N. Granda, Class. Quantum Grav. {\bf 28}, 025006 (2011).

\bibitem{Staro2016} A. A. Starobinsky, S. V. Sushkov, M. S. Volkov,  JCAP {\bf 1606}, 007 (2016).

\bibitem{Germani2010}
C. Germani and A. kehagias, Phys. Rev. Lett. {\bf 105}, 011302 (2010);
S. V. Sushkov, Phys. Rev. D {\bf 85}, 123520 (2012);
E. N. Saridakis and S. V. Sushkov, Phys. Rev. D {\bf 81}, 083510 (2010).


\bibitem{Bardeen1980}
J. M. Bardeen, Phys. Rev. D {\bf 22}, 1882 (1980).

\bibitem{Harrison1967}
E. R. Harrison, Rev. Mod. Phys. {\bf 39}, 862 (1967).

\bibitem{Hartle:1983ai}
  J.~B.~Hartle, S.~W.~Hawking,
  Phys.\ Rev.\ D {\bf28}, 2960 (1983).
\bibitem{Vilenkin:1984wp}
  A.~Vilenkin,
  Phys.\ Rev.\ D {\bf30}, 509 (1984).

\end{thebibliography}
\end{document}